\documentclass[preprint]{revtex4}
\usepackage{graphicx}

\begin{document}

\title{$Su(1,2)$ Algebraic Structure of the $XYZ$ Antiferromagnetic Model in Linear Spin-Wave Frame}
\author{Shuo Jin$^1$\footnote{Electronic
address: jinshuo@buaa.edu.cn},
 Bing-Hao Xie$^2$,
Hong-Biao Zhang$^3$, Jing-Min Hou$^4$} \affiliation{$^1$Department
of Physics, School of Science, Beihang University, Beijing,
100083,
P. R. China\\
$^2$Beijing Information Technology Institute, Beijing, 100101, P.
R. China\\
$^3$Institute of Theoretical Physics, Northeast Normal University,
Changchun, 130024, P.R.China\\
$^4$Department of physics, Southeast University, Nanjing, 210096,
P. R. China}
\begin{abstract}The $XYZ$ antiferromagnetic model in linear
spin-wave frame is shown explicitly to have an $su(1,2)$ algebraic
structure: the Hamiltonian can be written as a linear function of
the $su(1,2)$ algebra generators. Based on it, the energy
eigenvalues are obtained by making use of the similar
transformations, and the algebraic diagonalization method is
investigated. Some numerical solutions are given, and the results
indicate that only one group solution could be accepted in
physics.
\\
PACS number(s): %03.65.-w,
03.65.Fd,
%03.75.Fi,
05.30.Jp
\end{abstract}
\maketitle
\section{introduction}
%Lie algebraic methods have been useful in the study of problems in
%physics ever since Lie algebras were introduced by Sophus Lie at
%the end of the 19th century, especially after the development of
%quantum mechanics in the first part of the 20th century
%\cite{3gilmore3,3chengjq3,3humphreys3,3wybourne}. It is a powerful
%procedure for solving energy eigenvalue problems in some cases,
%such as in hydrogen atom system \cite{3mehra} and so on.
%Especially, for the physical systems with certain dynamical group
%symmetry, Lie algebraic methods are very useful in the
%diagonalization of the Hamiltonian
%\cite{3solomon,3amico,3montorsi,3pan,3xbh,3wang1,3wang2}, which is
%an important procedure in solving the many-body problem in second
%quantization frame.

As is well known, the Heisenberg model is a simple but realistic
and extensively studied solid-state system \cite{3hammar,3eggert}.
According to the sign of the interaction intensity $J$, the model
can be classified as the ferromagnetic type and the
antiferromagnetic one. Based on the interaction intensity along
the different space directions, the model can be labelled as the
$XXX$, $XXZ$, and $XYZ$ ones. Recently, it has been found that the
Heisenberg interaction is not localized in spin system, and it can
be realized in quantum dots \cite{3loss,3burkard}, nuclear spins
\cite{3kane}, cavity QED \cite{3imamoglu,3zhengsb}. Thus, the
study of this basic model is of interests and wide applications in
physical fields. In the investigation of these models, it is a
basic task to get the exact solutions
\cite{jin1,zhang1,pan1,pan2,birman1}, i.e., to diagonalize the
Hamiltonian. So far, only some special Heisenberg models can be
exactly solved, such as $XXX$ antiferromagnetic model
\cite{bethe}. In general, the linear spin-wave
\cite{3andersonpw,3kubo,callaway} approximation is widely applied
when we study the Heisenberg models. It has been known that the
$XXZ$ model in linear spin-wave frame can be diagonalized by
coherent state operators \cite{3xbh,3zwm} of $su(1,1)$ algebra.
However, for the $XYZ$ model in this frame, the method of the
coherent states does not work. Therefore, it is necessary for us
to develop another algebraic diagonalization method to get the
energy spectrum by using the algebraic structure which the model
has.

In this Letter, we review the $su(1,1)$ coherent states of the
$XXZ$ antiferromagnetic model in linear spin-wave frame. Then, the
$XYZ$ antiferromagnetic model in linear spin-wave frame can be
written as the generators of $su(1,2)$ algebra. At last, the
energy eigenvalues are obtained in terms of the algebraic
diagonalization method, and some numerical solutions are given and
discussed.
%Some general applications of this diagonalization method are analyzed at last.

%\section{The model}

\section{$XXZ$ antiferromagnetic model and $su(1,1)$ coherent states}
The Hamiltonian of the $XXZ$ antiferromagnetic model reads:
\begin{equation}
\label{hxxz} H_{XXZ}=-J\sum_{<i,j>}(S_i^xS_j^x+S_i^yS_j^y+\eta
S_i^zS_j^z)\;\;\; (J<0),
\end{equation}
where the notation $<i,j>$ denote the nearest neighbor bonds.
Starting from the two-sublattice model and Holstein-Primakoff
transformation \cite{3holstein}:
\begin{eqnarray}
\label{h-pt}
S_a^z&=&-s+a^{\dag}a,\;\;\;\;\;\;\;\;S_b^z=s-b^{\dag}b,\nonumber\\
S_a^{\dag}&=&(2s)^{\frac{1}{2}}(1-a^{\dag}a/2s)^{\frac{1}{2}}a,\;\;S_a^{-}=(S_a^{\dag})^{\dag},\nonumber\\
S_b^{\dag}&=&(2s)^{\frac{1}{2}}b^{\dag}(1-b^{\dag}b/2s)^{\frac{1}{2}},\;\;S_b^{-}=(S_b^{\dag})^{\dag},
\end{eqnarray}
where $a^+$ and $a$, ($b^+$, $b$) can be regarded as the creation
and annihilation operators of boson on sublattice A (sublattice
B), respectively, but the particle numbers $n_a=a^+a$, $n_b=b^+b$
can't excel $2s$, respectively. Because in low temperature and low
excitation condition, $<a^{\dag}a>,<b^{\dag}b>\ll s$, the
non-linear interaction in Eq. (\ref{h-pt}) could  be reasonable
ignored \cite{kittel}. Based on it, transferring
%\begin{eqnarray}
%\label{lhp}
%&&S_a^{\dag}=(2s)^{\frac{1}{2}}a,\ \ \ \ S_b^{\dag}=(2s)^{\frac{1}{2}}b^{\dag},\nonumber\\
%&&S_a^{-}=(2s)^{\frac{1}{2}}a^{\dag}\ \
 %\ \ S_b^{-}=(2s)^{\frac{1}{2}}b,
%\end{eqnarray}
the operators into momentum space,
%\begin{eqnarray}
%\label{moment} &&a_i=N^{-\frac{1}{2}}\sum_{{\bf k}}e^{i{\bf
%k}\cdot {\bf R}_i}a_{{\bf k}},\ \ \ \
%a_i^{\dag}=N^{-\frac{1}{2}}\sum_{{\bf k}}e^{-i{\bf
%k}\cdot {\bf R}_i}a_{{\bf k}}^{\dag},\nonumber\\
%&&b_j=N^{-\frac{1}{2}}\sum_{{\bf k}}e^{-i{\bf k}\cdot {\bf
%R}_j}b_{{\bf k}},\ \ \ \ b_j^{\dag}=N^{-\frac{1}{2}}\sum_{{\bf
%k}}e^{i{\bf k}\cdot {\bf R}_j}b_{{\bf k}}^{\dag},\nonumber\\
%\end{eqnarray}
we obtain
\begin{eqnarray}
\label{hsw xxz}
H_{XXZ}&=&-2ZsJ(Ns\eta-\sum_{\bf k}H_{\bf k}),\\
H_{\bf k} &=& \eta (a _{\bf k}^{\dag}a_{\bf k} +b_{\bf
k}^{\dag}b_{\bf k})+\gamma_{\bf k}(a_{\bf k}b_{\bf
k}+a^{\dag}_{\bf k}b^{\dag}_{\bf k}).
\end{eqnarray}
Here
\begin{equation}
\gamma_{\bf k}=Z^{-1}\sum_{\bf R} e^{i{\bf k}\cdot \bf R}
=\gamma_{-\bf k},
\end{equation}
in which $\bf R$ is a vector connecting an atom with its nearest
neighbor, and the sum runs over the $Z$ nearest neighbors. $2N$ is
the total number of the lattices. $H_{\bf k}$ can be expressed as
the linear combination of the $su(1,1)$ algebra generators in the
form
\begin{eqnarray}
\label{hk11} H_{\bf{k}}=2\eta E_{z}^{\bf k}+\gamma_{\bf
k}(E_{+}^{\bf k}+E_{-}^{\bf k}),
\end{eqnarray}
with
\begin{equation}
\label{generators11} E_{+}^{\bf k} =a^+_{\bf k}b^+_{\bf k},\;\;
E_{-}^{\bf k} =a_{\bf k}b_{\bf k},\;\; E_{z}^{\bf k}
=\frac{1}{2}(n^a_{\bf k}+n^b_{\bf k}+1),
\end{equation}
which obey the commutation relations of $su(1,1)$ Lie algebra:
\begin{equation}
\label{cr11} [E^{\bf k}_{+} , E^{\bf k}_{-}]=-2E^{\bf k}_z, \;\;
[E^{\bf k}_{z},E^{\bf k}_{\pm}]=\pm E^{\bf k}_{\pm}.
\end{equation}
By introducing an $su(1,1)$ displacement operator
\begin{equation}
\label{wxxz11} W(\xi_{\bf k})=\exp(\xi _{\bf k}E_{+}^{\bf k}-\xi
_{\bf k}^*E_{-}^{\bf k})
\end{equation}
with the coherent parameter $\xi_{\bf k}=r e^{i\theta}$,
%satisfying the identities:
%\begin{eqnarray}
%&&W_{\bf k}^{-1}(r)E^{\bf k}_{\pm}W_{\bf k}(r)\nonumber\\
%&&=E^{\bf k}_{z}e^{\pm i\theta}\sinh2r+E^{\bf k}_{\pm}\cosh^{2}r
%+E^{\bf k}_{\mp}e^{\mp 2i\theta}\sinh^2r,\nonumber\\
%\end{eqnarray}
%\begin{eqnarray}
%&&W_{\bf k}^{-1}(r)E^{\bf k}_{z}W_{\bf k}(r)\nonumber\\&&=E^{\bf
%k}_{z}\cosh2r +\frac{1}{2}E^{\bf k}_{+}e^{i\theta}\sinh2r +
%\frac{1}{2}E^{\bf k}_{-}e^{-i\theta}\sinh2r,\nonumber\\
%\end{eqnarray}
then we have
\begin{eqnarray}
\label{whwxxz} W^{-1}(\xi_{\bf k})H_{\bf k}W(\xi_{\bf k}) = \alpha
E^{\bf k}_{z}+(\beta E^{\bf k}_{+} +\beta ^* E^{\bf k}_{-}),
\end{eqnarray}
where
\begin{equation}
\label{alpha} \alpha=2\eta \cosh 2r+\gamma_{\bf
k}(e^{i\theta}+e^{-i\theta})\sinh2r,
\end{equation}
\begin{equation}
\label{beta} \beta=\eta e^{i\theta}\sinh2r+\gamma_{\bf
k}(\cosh^2r+e^{2i\theta}\sinh^2r).
\end{equation}
In order to diagonalize $H_{\bf{k}}$, the coefficient before the
non-Cartan generators $E^{\bf k}_{+}$ and $E^{\bf k}_{-}$ of Lie
algebra, $\beta$, should be chosen to zero (here we set $\theta=0$
for simplicity) and this leads to
\begin{equation}
\tanh2r=-\frac{\gamma_{\bf k}}{\eta},\;\;\;\; \alpha=2\sqrt{\eta
^2 -\gamma^2_{\bf k}}.
\end{equation}
So if we denote
\begin{equation}
|\xi_{\bf k}>=W(\xi_{\bf k})|vac\rangle,
\end{equation}
then one has
\begin{equation}
\label{enhxxz} H_{\bf k}|\xi_{\bf k}>=(n_a+n_b+1)\epsilon_{\bf
k}|\xi_{\bf k}>,
\end{equation}
\begin{equation}
\epsilon_{\bf k}=2ZJs\sqrt{\eta^2-\gamma_{\bf k}^2},
\end{equation}
where requiring $|\eta|\geq|\gamma_{\bf k}|$. The diagonalization
of the $XXZ$ antiferromagnetic model in linear spin-wave frame has
turned out to be a direct product of $su(1,1)$ coherent states
$\otimes_{\bf k} |\xi_{\bf k} \rangle$. One can see $\epsilon_{\bf
k}$ is the quantum of antiferromagnetic spin-wave, i.e. the
dispersion relation. From Eq. (\ref{enhxxz}), it is known that for
any {\bf k}, there exist two branches of degenerate
antiferromagnetic spin-wave which the quasi-particle numbers are
described by $n_a$ and $n_b$, respectively.

\section{$XYZ$ antiferromagnetic model in linear spin-wave frame with the $su(1,2)$ algebraic structure}

Owing to the different interaction intensity along the different
space directions, in general, the Hamiltonian of the $XYZ$
antiferromagnetic model is described by
\begin{eqnarray}
\label{hxyz} H_{XYZ}&=&-J\sum_{<i,j>}(\eta_{x} S_i^xS_j^x+\eta_{y}
S_i^yS_j^y+ S_i^zS_j^z)\nonumber\\
&&(J<0,\;\;\eta_{x},\eta_{y}>0),
\end{eqnarray}
where we have set $\eta_{z}=1$. Similar to the former case of the
$XXZ$ antiferromagnetic model, the $XYZ$ antiferromagnetic model
in linear spin-wave frame is given by the Hamiltonian:
\begin{eqnarray}
H_{XYZ}&=&2ZsJ[Ns-(\sum_{k}{\cal H}_{\bf{k}}-1)],\\
\label{hkxyz}
{\cal H}_{\bf{k}}&=&a _{\bf k}^{\dag}a_{\bf k} +b_{\bf k}^{\dag}b_{\bf k}\nonumber\\
&&+ \upsilon_{\bf k}(a_{\bf k}b_{\bf {-k}}^{\dag}+ a_{\bf
k}^{\dag}b_{\bf{-k}})+{\rho}_{\bf k}(a_{\bf
k}b_{\bf k}+a^{\dag}_{\bf k}b^{\dag}_{\bf k}),\nonumber\\
\end{eqnarray}
with
\begin{equation}
\upsilon_{\bf k}=\frac {\eta_{x}-\eta_{y}}{2}\gamma_{\bf k},\;\;
{\rho}_{\bf k}= \frac {\eta_{x}+\eta_{y}}{2}\gamma_{\bf k}.
\end{equation}
If we choose
\begin{eqnarray}
\label{esu12} & & I_{+}^{\bf k} =a^+_{\bf k}b_{-\bf
k},\;\;I_{-}^{\bf k} =a_{\bf k}b^{+}_{-\bf k},\;\;U_{+}^{\bf k}
=a_{\bf k}b_{\bf k},
\nonumber \\
& & V_{+}^{\bf k} =b^+_{\bf k}b^+_{-\bf k},\;\;V_{-}^{\bf k}
=b_{\bf k}b_{-\bf k},\;\;U_{-}^{\bf k} =a^+_{\bf k}b^+_{\bf k},
\nonumber \\
& & I_{3}^{\bf k} =\frac{1}{2}(n^a_{\bf k}-n^b_{-\bf
k}),\nonumber\\
&&I_{8}^{\bf k} =-\frac{1}{3} (n^a_{\bf k}+n^b_{-\bf k}+2n^b_{\bf
k}+2),
\end{eqnarray}
then, one can see they obey the commutation relations of $su(1,2)$
Lie algebra (here we omit the momentum sign ${\bf k}$):
\begin{eqnarray}
\label{re12} &&[I_{3},I_{\pm}]=\pm {I_\pm},\;[I_{+},I_{-}]=2I_{3},
[I_{8},I_{\alpha}]=0,(\alpha=\pm,3)\nonumber\\
&&[I_{3},U_{\pm}]=\mp\frac{1}{2}U_{\pm},[I_{8},U_{\pm}]=\pm
U_{\pm},[U_{+},U_{-}]=I_{3}-\frac{3}{2}I_{8},\nonumber\\
&&[I_{3},V_{\pm}]=\mp\frac{1}{2}V_{\pm}, [I_{8},V_{\pm}]=\mp V
_{\pm},[V_{+},V_{-}]=I_{3}+\frac{3}{2}I_{8},\nonumber\\
&&[I_{\pm},U_{\pm}]=\mp V_{\mp},[U_{\pm},V_{\pm}]=\pm
I_{\mp},[I_{\pm},V_{\pm}]=\pm U_{\mp},\nonumber\\
&&[I_{\pm},U_{\mp}]=[I_{\pm},V_{\mp}]=[U_{\pm},V_{\mp}]=0.
\end{eqnarray}
From Eqs. (\ref{hkxyz}) and (\ref{esu12}), ${\cal H}_{\bf{k}}$ can
be expressed as the linear combination of six generators of Lie
algebra $su(1,2)$ and processes $su(1,2)$ algebraic structure,
i.e.,
\begin{eqnarray}
\label{hsu12} {\cal H}_{\bf{k}}=I_{3}^{\bf
k}-\frac{3}{2}I_{8}^{\bf k}+{\rho}_{\bf k}(I_{+}^{\bf
k}+I_{-}^{\bf k})+ \upsilon_{\bf k}(U_{+}^ {\bf{k}}+U_{-}^{\bf
k}).
\end{eqnarray}

\section{the diagnolization and the eigenvalues}
If we set the general linear combination form of Lie algebra
$su(1,2)$ as
\begin{eqnarray}
\label{h0xyz}
H_0&=&aI_{+}+bI_{-}+cU_{+}+dU_{-}\nonumber\\
&&+eV_{+}+fV_{-}+gI_{3}+hI_{8},
\end{eqnarray}
for ${\cal H}_{\bf{k}}$ (\ref{hsu12}), the coefficients in
Eq. (\ref{h0xyz}) are:
\begin{eqnarray}
\label{abcdxyz}
&&a=b={\rho}_{\bf k},\;\;\;c=d=\upsilon_{\bf k},\;\;\;e=f=0,\nonumber\\
&&g=1,\;\;\;\;\;h=-\frac{3}{2}.
\end{eqnarray}

Until now, the coherent state operator as Eq. (\ref{wxxz11}) has
not been found for the $XYZ$ antiferromagnetic model in linear
spin-wave frame. But following the standard Lie algebraic theory
\cite{3zwm,3wsj6,3gilmore3,3chengjq3,3humphreys3}, if $H_0$ is a
linear function of the generators of a compact semi-simple Lie
group, it can be transformed into a linear combination of the
Cartan operators of the corresponding Lie algebra by
\begin{eqnarray}
\label{h1xyz} {\cal H}_1={\cal W}{\cal H}_0{\cal W}^{-}.
\end{eqnarray}
Here ${\cal W}=\prod_{i=1}^Nexp(x_iA_i)$ is an element of the
group and ${\cal W}^-$ denotes the inverse of ${\cal W}$, in which
{$A_i$} ($i=1,...,N$) is a basis set in Cartan standard form of
the semi-simple Lie algebra, and $x_i$ can be set to zero if the
corresponding $A_i$ is a Cartan operator. By choosing
\begin{eqnarray}
\label{wxyz} {\cal
W}&=&exp(x_{31}V_+)exp(x_{21}I_-)exp(x_{32}U_-)\nonumber\\
&&exp(x_{12}I_+)exp(x_{23}U_+)exp(x_{13}V_-),
\end{eqnarray}
and letting the coefficients of the non-Cartan operators vanish,
while substituting Eqs. (\ref{h0xyz})(\ref{abcdxyz})(\ref{wxyz})
into the right-hand side of Eq. (\ref{h1xyz}), we get a complete
set of algebraic equations of $x_{ij}$ after lengthy computation:
\begin{equation} \label{ab3} \left\{
\begin{array}{l}
-(h+\frac{1}{2}g)x_{13}+(a+dx_{13})x_{23}=0\\
c+bx_{13}+(\frac{1}{2}g-h)x_{23}+dx^2_{23}=0\\
a+dx_{13}-(g+dx_{23})x_{12}-bx^2_{12}=0,
\end{array}
\right.
\end{equation}
and the Hamiltonian after the transformation of ${\cal W}$ becomes
diagonal:
\begin{eqnarray}
\label{whw12} {\cal H}_1&=&{\cal W}{\cal H}_{0}{\cal
W}^{-}\nonumber\\
&=&(g+dx_{23}+2bx_{12})I_{3}+(h-\frac{3}{2}dx_{23})I_{8}.
\end{eqnarray}
One can see that although operator ${\cal W}$ is not unitary, the
similar transformation (\ref{h1xyz}) guarantees that the
eigenvalues of ${\cal H}_0$ equal those of ${\cal H}_1$. This is
acceptable for we are only concerned with the eigenvalues. If the
total particle number ${\cal N}=n^a_{\bf k}+n^b_{-\bf k}+n^b_{\bf
k}$ ($n^a_{\bf k}=a_{\bf k}^{\dag}a_{\bf k},\ n^b_{-\bf k}=b_{-\bf
k}^{\dag}b_{-\bf k},\ n^b_{\bf k}=b_{\bf k}^{\dag}b_{\bf k}$),
from Eq. (\ref{re12}), $[\Gamma,{\cal N}]=0\ \
({\Gamma}=I_\pm,V_\pm,U_\pm,I_3,I_8)$ holds. Hence, supposing the
common eigenstates of the Cartan generators $I_3$ and $I_8$ of Lie
algebra $su(1,2)$ are the Fock states $\mid n^a_{\bf k},n^b_{-\bf
k},n^b_{\bf k}>$, i.e., for the commutative set $\{I_3,I_8,{\cal
N}\}$ there exist:
\begin{eqnarray}
\label{fock12} &&I_3\mid n^a_{\bf k},n^b_{-\bf k},n^b_{\bf k}>
=\frac{1}{2}(n^a_{\bf k}-n^b_{-\bf k})\mid n^a_{\bf k},n^b_{-\bf k},n^b_{\bf k}>,\nonumber\\
&&I_8\mid n^a_{\bf k},n^b_{-\bf k},n^b_{\bf k}>\nonumber\\
&&=-\frac{1}{3}(n^a_{\bf k}+n^b_{-\bf k}+2n^b_{\bf
k}+2)\mid n^a_{\bf k},n^b_{-\bf k},n^b_{\bf k}>,\nonumber\\
&&{\cal N}\mid n^a_{\bf k},n^b_{-\bf k},n^b_{\bf k}>\nonumber\\
&&=(n^a_{\bf k}+n^b_{-\bf k}+n^b_{\bf k})\mid n^a_{\bf
k},n^b_{-\bf k},n^b_{\bf k}>.
\end{eqnarray}
From Eqs. (\ref{whw12})(\ref{fock12}) it follows the eigenvalue of
the Hamiltonian (\ref{hsu12}):
\begin{eqnarray}
%E&=&\frac{1}{2}(g+dx_{23}+2bx_{12})(n^a_{\bf k}-n^b_{-\bf
%k})\nonumber\\
%&&-\frac{1}{3}(h-\frac{3}{2}dx_{23})(n^a_{\bf k}+n^b_{-\bf
%k}+2n^b_{\bf k}+2),
E&=&\omega^{a}_{\bf k}n^a_{\bf k}+\omega^{b}_{\bf k}n^b_{\bf
k}+\omega^{b}_{-\bf k}n^b_{-\bf
k}+\omega^{E}_{\bf k},\nonumber\\
\omega^{a}_{\bf
k}&=&(\frac{1}{2}g-\frac{1}{3}h+bx_{12}+dx_{23}),\nonumber\\
\omega^{b}_{\bf k}&=&(-\frac{2}{3}h+dx_{23}),\nonumber\\
\omega^{b}_{-\bf k}&=&(\frac{1}{2}g-\frac{1}{3}h-bx_{12}),\nonumber\\
\omega^{E}_{\bf k}&=&-\frac{2}{3}h+dx_{23},
\end{eqnarray}
where the coefficients $b,d,g,h,$ are given in Eq. (\ref{abcdxyz})
and $x_{ij}$ can be obtained by solving Eq. (\ref{ab3}), and
$\omega^{a}_{\bf k},\omega^{b}_{\bf k},\omega^{b}_{-\bf k}$ are
the energies of the three different magnons respectively. In fact,
the order of the operators in ${\cal W}$ can be chosen
arbitrarily, but the coefficients $x_i$ are strongly dependent on
the order. Although any specified order has a solution, a properly
chosen order can simplify the procedure to get the $x_i$. In
general, for the Hamiltonian with $su(n)$ (whose Cartan operators
are $A_{ii}=b_i^+b_i$) or the isomorphic algebra
$su(p,q)\;\;(p+q=n)$ structure, the transformation operator ${\cal
W}$ can be chosen as
\begin{eqnarray}
\label{wusual} {\cal
W}&=&exp(x_{N1}A_{N1})exp(x_{N2}A_{N2})...\nonumber\\
&&exp(x_{2N}A_{2N})exp(x_{1N}A_{1N}),
\end{eqnarray}
where the order of the operators of $exp(x_{ij}A_{ij})\;(i\neq j)$
is arranged according to the roots of $A_{ij}$ in a decreasing
way. For example, the root of $A_{N1}$ is highest, and that of
$A_{1N}$ is lowest. As a rule of our choice, the right or left
$A_{ij}$ in Eq. (\ref{wusual}) is the one that is missing in the
Hamiltonian ${\cal H}_0$; the middle operators sequence forms a
circle root diagram. With this specification, in our experience
the coefficients $x_{ij}$ are relatively easy to work out. We also
choose the form of ${\cal W}$ in Eq. (\ref{wxyz}) as
$exp(x_{13}V_-)exp(x_{12}I_+)exp(x_{23}U_+)
exp(x_{21}I_-)exp(x_{32}U_-)exp(x_{31}V_+)$, and it can be proofed
they lead to the same eigenvalues.

\section{numerical solutions}
From the Eq. (\ref{ab3}), maybe there have several sets of
solutions, which consist the complete set. But the different sets
of the slutions cannot be the eigenstate of the Hamiltonian
(\ref{hkxyz}) or (\ref{hsu12}) simultaneously. Only those who
possess the physical meaning is the solutions we need. In order to
illustrate it, we consider a concrete example of the Hamiltonian
(\ref{hkxyz}) or (\ref{hsu12}) with
$\eta_{x}=0.8,\eta_{y}=0.5,\gamma_{\bf k}=1$. Then $v_{\bf
k}=0.15,\rho_{\bf k}=0.65$. Using maple, one can show that there
are six sets solutions of Eq. (\ref{ab3}), in which only one set
$x_{13}=-0.6018692595e-1,x_{23}=0.9389946768e-1,x_{12}=0.4827143955$
leads to the positive energy $\omega^{a}_{\bf
k}=1.327849277,\omega^{b}_{\bf k}=1.014084920,\omega^{b}_{-\bf
k}=0.6862356430$ of the magnons. It is clear that only this
solution is the accepted in physics. Other five sets solutions
(with negative energy) are non-physical. This procedure is easily
to be taken in solving the Eq. (\ref{ab3}) and Eq. (\ref{hsu12}).

\section{Conclusion and remarks}

In conclusion, the eigenvalue problems of the $XYZ$
antiferromagnetic model in linear spin-wave frame is solved from
an algebraic point of view. To use this algebraic diagonalization
method, first, it is needed to find the algebraic structure of the
Hamiltonian and manage to write the Hamiltonian into a linear
combination form of the algebraic generators just as Eq.
(\ref{hsu12}). Second, according the particular structure of some
Lie algebra, one may looking for the transformation operator. The
key is that we can let the coefficients of the non-Cartan
operators vanish successfully and get the solvable equations of
the parameters through the transformation. Some numerical
solutions check our diagonalization method, whose advantage is
that the eigenvalues of ${\cal H}_0$ equal those of ${\cal H}_1$
although the Hamiltonian changes. Due to the different interaction
intensity along the different space directions for the Heisenberg
model, the complication in physical model leads to the enlargement
of the algebra structure, such as the $XXZ$ case to the $XYZ$ case
corresponds to the $su(1,1)$ algebra to the $su(1,2)$ one. Of
course, the change of the algebra structure brings the different
method in diagonalizing the Hamiltonian. It is reasonable to
believe that more useful physical applications of the algebraic
diagonalization method should be found. It may be possible to
extend this case to higher-rank Lie algebras.

\begin{acknowledgments} This work is in part supported by the National Science Foundation
of China under Grant No. 10447103, Education Department of Beijing
Province and Beihang University.
\end{acknowledgments}

\end{document}